\documentclass[journal=apchd5,manuscript=article]{achemso}
\setkeys{acs}{maxauthors = 500} 

\usepackage{chemformula} 
\usepackage[T1]{fontenc} 

\usepackage{pgfplots}
\usepackage{epstopdf}
\usepackage{graphicx}
\graphicspath{{./Figures/}}

\usepackage[nolist]{acronym} 
\usepackage{siunitx} 
\usepackage{dsfont} 
\usepackage{xcolor, soul} 
\usepackage{placeins} 

\author{Gökhan Kara}
\affiliation[TNI]
{Transport at Nanoscale Interfaces Laboratory, Empa – Swiss Federal Laboratories for Materials Science and Technology, CH-8600 Dübendorf, Switzerland}
\author{Lorenzo J. A. Ferraresi}
\affiliation[TNI]
{Transport at Nanoscale Interfaces Laboratory, Empa – Swiss Federal Laboratories for Materials Science and Technology, CH-8600 Dübendorf, Switzerland}
\alsoaffiliation[ETHC]
{Department of Chemistry and Applied Biosciences, ETH – Swiss Federal Institute of Technology Zurich, CH-8093 Zurich, Switzerland}
\author{Dmitry N. Dirin}
\affiliation[ETHC]
{Department of Chemistry and Applied Biosciences, ETH – Swiss Federal Institute of Technology Zurich, CH-8093 Zurich, Switzerland}
\alsoaffiliation[TFP]
{Laboratory for Thin Films and Photovoltaics, Empa – Swiss Federal Laboratories for Materials Science and Technology, CH-8600 Dübendorf, Switzerland}
\author{Roman Furrer}
\affiliation[TNI]
{Transport at Nanoscale Interfaces Laboratory, Empa – Swiss Federal Laboratories for Materials Science and Technology, CH-8600 Dübendorf, Switzerland}
\author{Maksym V. Kovalenko}
\affiliation[ETHC]
{Department of Chemistry and Applied Biosciences, ETH – Swiss Federal Institute of Technology Zurich, CH-8093 Zurich, Switzerland}
\alsoaffiliation[TFP]
{Laboratory for Thin Films and Photovoltaics, Empa – Swiss Federal Laboratories for Materials Science and Technology, CH-8600 Dübendorf, Switzerland}
\author{Michel Calame}
\affiliation[TNI]
{Transport at Nanoscale Interfaces Laboratory, Empa – Swiss Federal Laboratories for Materials Science and Technology, CH-8600 Dübendorf, Switzerland}
\alsoaffiliation[UB]
{Department of Physics and Swiss Nanoscience Institute, University of Basel , CH-4056 Basel, Switzerland}
\author{Ivan Shorubalko}
\affiliation[TNI]
{Transport at Nanoscale Interfaces Laboratory, Empa – Swiss Federal Laboratories for Materials Science and Technology, CH-8600 Dübendorf, Switzerland}
\email{ivan.shorubalko@empa.ch}

\title[]
  {The Comparison of Colloidal PbS QD Photoconductors and Hybrid Phototransistors}

\keywords{colloidal quantum dots, graphene, phototransistors, photoconductors, infrared photodetectors, temperature dependence}

\begin{document}

\begin{acronym}
	\acro{NPC}{negative photocurrent}
	\acro{PPC}{positive photocurrent}
	\acro{cQD}{colloidal quantum dot}
	\acro{EDT}{ethane-1,2-dithiol}
	\acro{SWIR}{short wave infrared}
	\acro{IFP}{interdigitated fingers photodetector}
	\acro{HP}{hybrid photodetector}
	\acro{L}{channel length}
	\acro{W}{channel width}
	\acro{CVD}{chemical vapor deposition}
	\acro{CMOS}{complementary metal-oxide-semiconductor}
	\acro{LbL}{layer-by-layer}
	\acro{GFET}{graphene field effect transistor}
	\acro{FET}{field effect transistor}	
\end{acronym}


\begin{tocentry}

\includegraphics[width=\linewidth]{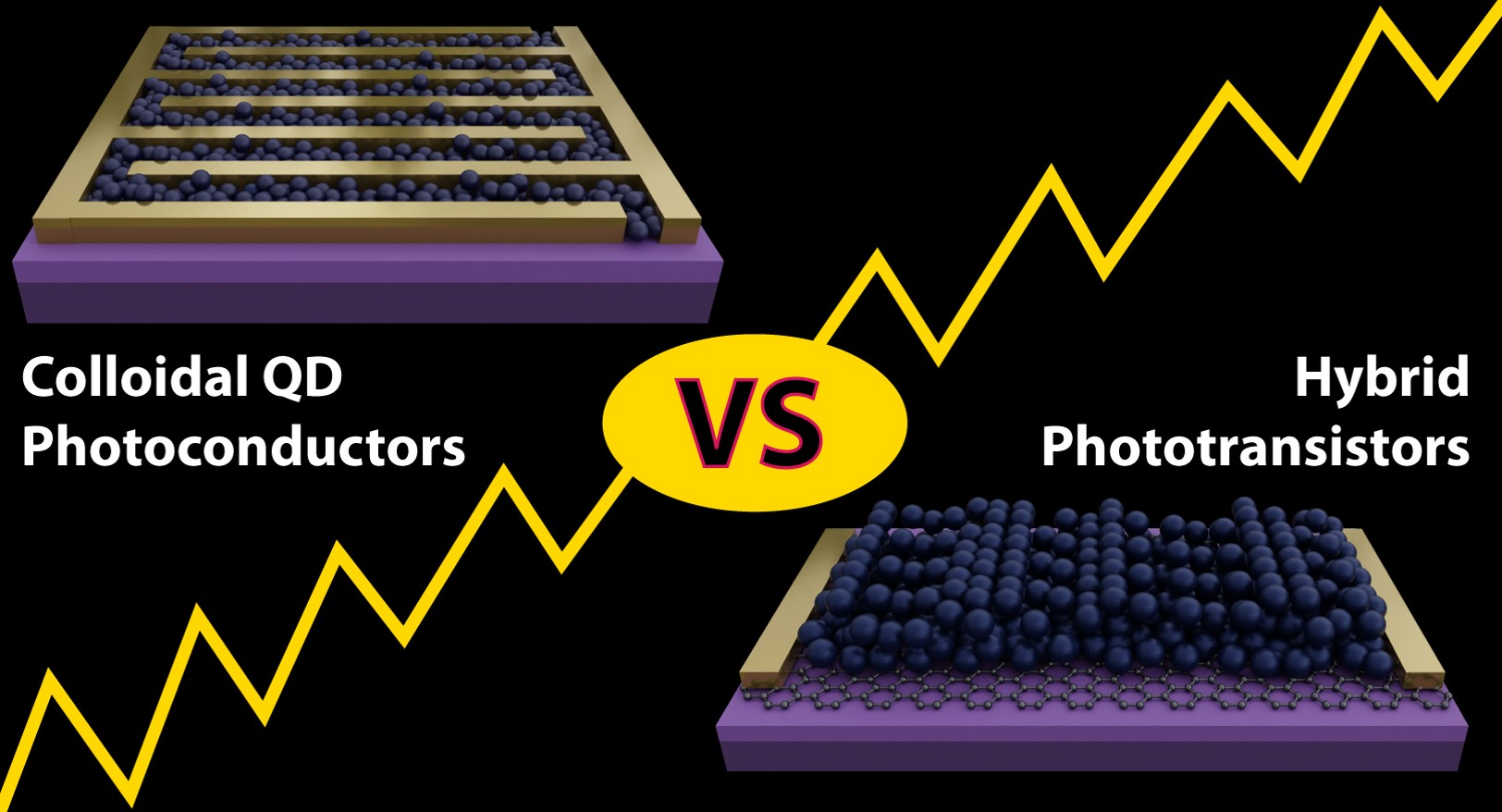}

\end{tocentry}

\newpage
\begin{abstract}
  The simplicity in the fabrication of photoconductors makes them a valuable choice to investigate optoelectronic properties of \acf{cQD} films. Lateral photoconductors generally require a large size, in the \unit{mm^2}, and are limited in operation speed due to the presence of trapping sites. In contrast, hybrid phototransistors are fabricated in the \unit{\um^2} scale and benefit from such trapping sites, allowing the measurement of low light levels in the \unit{nW \per cm^2}. The question, however, arises whether high responsivity values are required for the detection of low light levels or the compatible detectivity of photoconductors is sufficient. Here, we directly compare photoconductors and hybrid phototransistors with an identical \acs{EDT} treated PbS \acs{cQD} film. We highlight that a comparable $D^*$ is not enough for the purpose of measuring low light levels, as the resulting photocurrents need to be readily accessible. Furthermore, we also showcase temperature-activated photocurrent dynamics resulting in an \acf{NPC} effect. This \acs{NPC} simultaneously improves the frequency bandwidth and photocurrent, enabling operation speeds up to \qty{100}{kHz}.
\end{abstract}
\newpage

\section*{Introduction}

Advances in the research on \acfp{cQD} has led to remarkable developments for light-emitting devices (e.g., QLEDs), with a recent demonstration of electrically driven amplified spontaneous emission from a \acp{cQD} diode.\cite{Ahn2023} Likewise, \acsp{cQD} open up new applications in imaging beyond the silicon bandgap. PbS \acsp{cQD}, in particular, can be tuned across the \acf{SWIR} spectral region (1 to about \qty{2.5}{\um} in wavelength)\cite{Moreels2011} and are highly desired for biological imaging, night vision, or telecommunication.\cite{GarciadeArquer2021,Manna2023}

Photodetectors based on PbS \acs{cQD} are commonly fabricated as photoconductors, photodiodes, or phototransitors.\cite{Saran2016} Due to their ease of fabrication, lateral \acs{cQD} photoconductors were among the first devices studied and used to investigate \acs{cQD} film properties.\cite{Konstantatos2006} In those devices, the photoconducting material acts simultaneously as the photoabsorber and the conductive channel. The low charge carrier mobility of these disordered films (typically $<10^{-1}$ \unit{cm^2 \per Vs})\cite{GarciadeArquer2021}, results in large device structures of up to \unit{mm^2} to achieve accessible resistances in the range of $\sim$\unit{M\Omega}'s.\cite{Saran2016} Trapping sites slow down the charge transfer through the structure and limit the operation speed.\cite{Saran2016,Kahmann2020} This speed limitation was overcome by vertical photodiode architectures, which drastically reduce the charge extraction distance to a few hundred \unit{nm} and lead to response times down to \unit{ns}.\cite{Vafaie2021}

On the contrary, the hybrid phototransistor sacrifices operation speed for amplification of the photosignal. A \acs{cQD} layer on top of a 2D material, such as graphene, acts as a photogate.\cite{Konstantatos2012} The main benefit of this structure is the decoupling of the photoabsorbing layer (\acs{cQD}) and conducting channel (graphene). Trapped charges in the \acs{cQD} layer are now an advantage and give rise to a photogain, as they directly modulate the current in the graphene channel. The observed photogating effect enables the measurement of light intensities down to \unit{nW\per cm^2}.\cite{Konstantatos2012,Goossens2017}

Due to the inherent photogain, the phototransistor can reach responsivity values $R$ in the order of $10^6$ to $10^9$ \unit{A\per W}.\cite{Konstantatos2012,Turyanska2015,Ahn2020} However, this does not always translate in high detectivity values $D^*$, as graphene also supports a large dark current. In contrast, photoconductors promote moderate $R$ values in the range of $10^{-2}$ to \qty{1}{A\per W},\cite{Saran2016} but generally yield competitive $D^*$ merits. Due to their low dark currents, reported $D^*$ values are in the range between $10^9$ to $10^{12}$ \unit{Jones},\cite{Saran2016} and are comparable to the performance of hybrid phototransistors.\cite{Ahn2020,Grotevent2021,Kara2023} Comparing the performance of individual photodetectors is, thus, challenging due to the different architectures and, additionally, because of differently treated \acs{cQD} films even for the same material composition.

Here, we compare the lateral photoconductor and hybrid phototransistor, fabricated on the same sample. This allows for a direct comparison of both architectures based on the identical photoactive material solely. We highlight the advantages and limitations of each architecture, and emphasize the importance of a high responsivity $R$ despite a comparable detectivity $D^*$ value. For a technologically representative device area of $\sim$ \qty{20}{\um ^2}, the photoconductor reaches a photocurrent, $I_{ph}, of \sim$ \qty{100}{fA}, while the phototransistor achieves $I_{ph}\sim 100$ \unit{nA}. Since reducing pixel size is essential for high-resolution imaging, larger photocurrents are advantageous for practical pixel readout, thereby favoring high responsivity despite similar $D^*$. We further investigate the role of AC-modulated versus continuous illumination, which reveals distinct photocurrent dynamics in \acs{EDT}-treated PbS \acs{cQD} films. Temperature-dependent measurements additionally highlight regimes of \acf{NPC}, which we reason for the observed simultaneous improvement of frequency bandwidth and photocurrent in hybrid phototransistors.

\section*{Results and discussion}

\subsection*{Interdigitated Fingers Photodetector}

\begin{figure}[h!tb] 
	\centering
	\includegraphics[width=\linewidth]{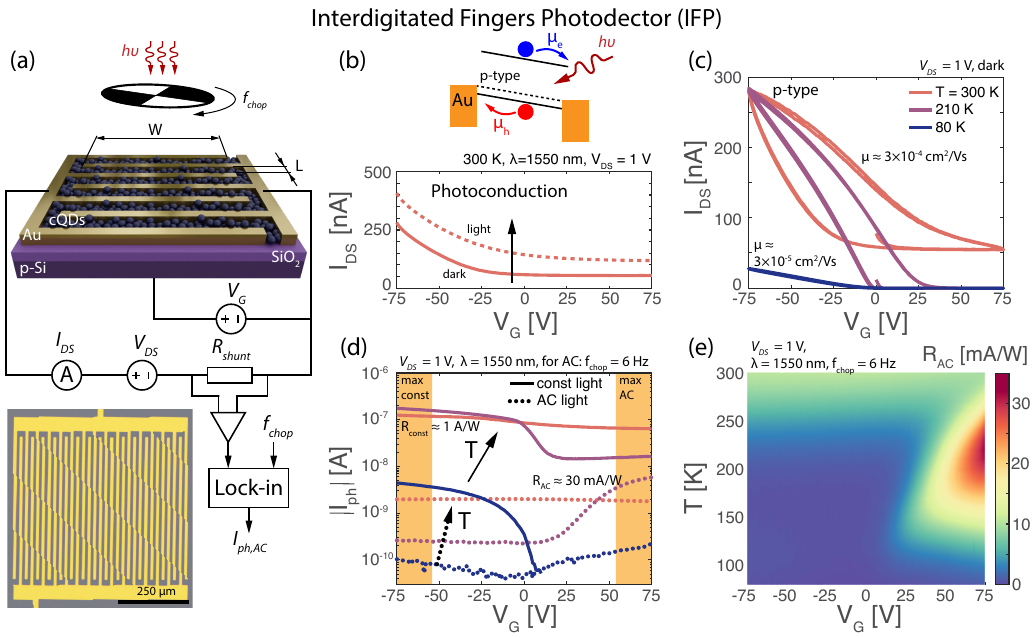}
	\caption{\Acf{IFP}. (a) \acs{IFP} consisting of 30 gaps with a \acf{L} of \qty{10}{\um} and a \acf{W} of \qty{500}{\um}, fabricated on a p-Si/SiO\textsubscript{2} substrate. A $\sim$ \qty{170}{nm} PbS \acf{cQD} film was spin-coated subsequently. A source-drain voltage $V_{DS}$ and a back-gate voltage $V_G$ were applied, while the source-drain current $I_{DS}$ was measured. A chopper modulated the incoming light at a frequency $f_{chop}$, causing a photocurrent $I_{ph, AC}$ on top of $I_{DS}$. $I_{ph, AC}$ was extracted by a lock-in technique over the voltage drop across a shunt resistor $R_{shunt}$. The inset shows the structure before \acs{cQD} deposition. (b) \acs{EDT} treated PbS \acs{cQD} film energy diagram. A p-type charge transport behavior follows from a higher hole mobility $\mu_h$ than an electron mobility $\mu_e$. The photoconduction of the \acs{IFP} is expressed as a vertical shift of the source-drain current upon illumination. (c) $V_G$-dependent $I_{DS}$ measured in the dark for three different temperatures. The curves show two loops (forward and backward sweeps). (d) Comparison of $V_G$-dependent photocurrents measured at constant illumination, $I_{ph,const} = I_{DS,light} - I_{DS,dark}$ (solid lines), or AC modulated, $I_{ph,AC}$ (dotted lines), at \qty{6}{Hz}. Curves are shown for the same temperatures as in (c). For simplicity, the constant illuminated curves are shown in the forward sweep direction only (e) AC modulated responsivity  $R_{AC}=I_{ph,AC}/P_{in}$, mapped over temperature $T$ (80 to \qty{300}{K}) and $V_G$. All the measurements upon light exposure were performed at the first excitonic peak (\qty{1550}{nm} wavelength) of the $\sim$\qty{6}{nm} \acsp{cQD}, and an irradiance ($\mathds{I}_{in}=P_{in}/A$) of \qty{120}{\mu W \per cm^2}. $V_{DS}$ was kept at \qty{1}{V} for all the measurements.}
	\label{fig:ch_TDependence_Figure1}
\end{figure}

Figure \ref{fig:ch_TDependence_Figure1} (a) shows the investigated \acf{IFP} structure. The detector consisted of gold fingers with a \acf{W} of \qty{500}{\um}, 30 gaps ($n_{gaps}$), and was spaced by \acf{L} of \qty{10}{\um}. In between these gold electrodes, a layer of PbS \acs{cQD} with a thickness $\sim$ \qty{170}{nm} was spin-coated layer-by-layer, replacing the native oleic acid ligands with \acf{EDT}. The device was characterized by applying a voltage $V_{DS}$ between the two electrode sides (source-drain) and measuring the current $I_{DS}$. Upon modulation of the incoming light with a frequency $f_{chop}$, $I_{DS}$ carried an additional AC photocurrent $I_{ph, AC}$, that was typically two to five orders of magnitude lower than the bias current itself (see Supporting Information). The AC photocurrent was extracted with a lock-in amplifier by measuring the voltage drop over a shunt resistance. A back gate voltage to the substrate, $V_G$, was used to also allow for modulation of the charger carrier density in the \acs{cQD} film.

The \acs{IFP} structure is a photoconductor, where the light absorber is simultaneously the conduction channel. In Figure \ref{fig:ch_TDependence_Figure1} (b), the energy diagram for such a typical \acs{EDT} ligand exchanged PbS \acs{cQD} photoconductor is drawn. Ligand exchange in \acs{cQD} films increases conductivity by decreasing the interparticle spacing and the energy barrier between the individual \acsp{cQD}.\cite{Liu2010a,Brown2014,Shcherbakov-Wu2020} An \acs{EDT} ligand exchange treatment, in particular, is commonly used to achieve hole transport layers (p-type). It is not the \acs{EDT} ligands, but the subsequent oxidation in air that creates predominant charge traps for electrons. This reduces the charge carrier mobility for electrons, $\mu_e$, giving the film a higher conductance for holes ($\mu_h > \mu_e$).\cite{Klem2008a,Balazs2014,Kahmann2020,Nagpal2011} Upon light modulation of the film, the current is increasing.

Figure \ref{fig:ch_TDependence_Figure1} (c) shows the investigated device's $V_G$-dependent dark current for three different temperatures. At \qty{300}{K}, the applied positive gate voltage was not able to fully deplete the film, which is expressed in a $\sim$\qty{50}{nA} offset current. Cooling the device down revealed the p-type character of the \acs{cQD} film. The threshold gate voltage between 0 to \qty{20}{V} was obscured by hysteresis effects. The dark current also got lower in the p-doped side of the channel ($V_G < 0$ \unit{V}). As the charge transport in \acs{cQD} films is generally described by hopping transport,\cite{Guyot-Sionnest2012} cooling caused a drop in the extracted field effect mobility from about $3\times10^{-4}$ to $3\times 10^{-5}$ \unit{cm^2 \per Vs}. All observed mobilities were in good agreement with commonly reported values ranging between $10^{-4}$ to $10^{-6}$ \unit{cm^2 \per Vs}.\cite{Klem2008a,Balazs2014,Speirs2016} The off state of the channel was about 3 \unit{G\Omega} (see Supporting Information) for temperatures below \qty{210}{K} and limited by gate leakage currents.

The comparison of photocurrents at a constant illumination,
\begin{equation}
	I_{ph,const} = I_{DS,light} - I_{DS,dark},
	\label{eq:ch_TDependence_IphDC}
\end{equation}
to the AC modulate light $I_{ph,AC}$ at \qty{6}{Hz}, are shown in Figure \ref{fig:ch_TDependence_Figure1} (d). $I_{DS,dark}$ and $I_{DS,light}$ are the source-drain current in the dark and at constant illumination, respectively (see Supporting Information). For simplicity, only a trace in the forward sweeping direction is shown. $I_{ph,const}$ was about constant at room temperature. At \qty{210}{K}, the photocurrent was slightly higher for negative $V_G$ compared to \qty{300}{K}. This is generally explained by competing mechanisms of higher charge carrier mobilities against shorter trapping times and can lead to an initial increase of the photocurrent if cooled down.\cite{Espevik2003,Konstantatos2007a,Konstantatos2008} For positive $V_G$, the photocurrent saturated for both 300 and \qty{210}{K}. A non-zero photocurrent in this regime is remarkable at \qty{210}{K} as the conducting channel is turning highly resistive. In contrast, the higher photocurrent at \qty{300}{K} for positive $V_G$ might be explained by the available thermal energy supporting hopping transport. When measured at a temperature of \qty{80}{K}, the photocurrent was quenched due to the reduced charge extraction length induced by low charge carrier mobilities.

The AC-modulated photocurrent showed a lower magnitude than the $I_{ph,const}$. As the lifetimes of photogenerated electron-hole (e-h) pairs are in the $\sim 1$ \unit{s} (see Supporting Information), a \qty{6}{Hz} modulation ($\sim0.2$ \unit{s}) did not allow to re-establish the dark current and resulted in $I_{ph,AC}<I_{ph,const}$. The temperature trend shows that reduced activation energy for hopping results in fewer charges being extracted to the electrodes. The photocurrent at a positive $V_G$ for the measurement at \qty{80}{K} was discarded for constant illumination because of the contribution from gate leakage to the dark current. Conversely, the AC-modulated photocurrent was extracted by a look-in technique and was thus unaffected by leakage currents.

The highest photocurrents for constant illumination (DC response) and the \qty{6}{Hz} modulated (AC response) were found at opposing sites for negative and positive $V_G$, respectively. This contradiction stresses the importance of charge carrier dynamics in \acs{cQD} films. It has already led to inconsistency if charge carrier mobilities are increasing (DC investigation)\cite{Liu2010a} or decreasing (AC investigation)\cite{Gilmore2017} with the nanocrystal size.\cite{Shcherbakov-Wu2020} Reported photodetectors do not always consider these photocurrent dynamics, making a performance cross-comparison more challenging than different device architectures already provoke. Transport in \acsp{cQD} can be described by hopping transport, where charge carriers jump between localized sites.\cite{Yu2004,Guyot-Sionnest2012,Shcherbakov-Wu2020} cQD films do have a rich trap state distribution caused by inhomogeneities in size, shape, and composition, as well as the surface of individual \acsp{cQD} themselves.\cite{Shcherbakov-Wu2020} Additionally, the \acs{cQD} film processing can lead to chemical residuals, oxidation, or formation of dimers (fused adjacent nanocrystals).\cite{Kahmann2020} The transport through such a variety of trap states directly alters the dynamic photoresponse.

The mismatch of constant to AC modulated photoresponse is also expressed in the magnitude of the responsivity values,
\begin{equation}
	R = \frac{I_{ph}}{P_{in}}.
\end{equation}
Here, $P_{in}$ is the area-dependent light power incident on the detector. $R_{const}\approx$ \qty{1}{A\per W} was reached with a constant illumination, leading to a
\begin{equation}
	EQE = \frac{\Phi_e}{\Phi_{ph}} = R\frac{E_{ph}}{e}
\end{equation}
of about 80\%. $\Phi_e$ and $\Phi_{ph}$ are the photogenerated e-h flux and the incoming photon flux respectively. $E_{ph}$ is the photon energy and $e$ is expressing the elementary charge. The achieved AC light modulated responsivity, $R_{AC}$, was about one order of magnitude lower instead. Figure \ref{fig:ch_TDependence_Figure1} (e) shows $R_{AC}$ mapped over gate voltage and temperature. The highest responsivity of $\sim 30$ \unit{mA/W} results in an $EQE$ of about 3\%. Following the trend of the photocurrent, the highest responsivity values were found around \qty{210}{K}, which emphasizes that the photocurrent dynamics strongly depend on temperature. The observed $EQEs$ indicate no photogain as all values were below 100\%. Comparable devices in the literature show responsivity values between \qty{20}{mA/W} to \qty{5}{A/W}.\cite{Saran2016} This results in $EQEs$ up to $\sim$ 1000\% and indicate that a photogain is possible in this architecture.

\subsection*{Hybrid Photodetector}

\begin{figure}[h!tb] 
	\centering
	\includegraphics[width=\linewidth]{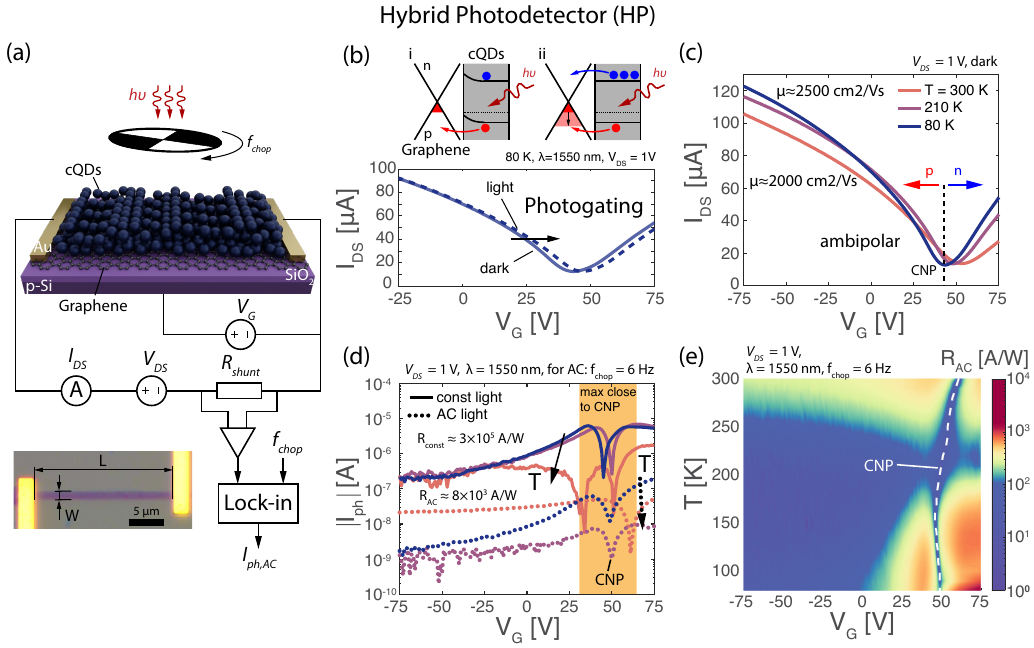}
	\caption{\Acf{HP}. (a) \acs{CVD} graphene with a channel of $L\times W= 20\times1$ \unit{\um ^2} was patterned on a p-Si/SiO\textsubscript{2} substrate. The measurements were performed as described above. Inset shows the graphene channel before the layer-by-layer spin-coating of PbS \acsp{cQD}. (b) Energy diagram for the graphene-\acs{cQD} interface. (i) The Schottky-like junction supports a higher hole than the electron-transfer rate across the graphene-\acs{cQD} film interface. This results in a negatively charged \acs{cQD} film, top-gating the graphene channel (ii). The photogating effect is expressed as a horizontal shift of the $I_{DS}-V_G$ curves upon light illumination. (c) $V_G$-dependent channel current $I_{DS}$, measured in dark and for three different temperatures. After cooling, the charge carrier mobility of the channel increased from 2000 to about \qty{2500}{cm^2 \per Vs}. On the left and right of the charge neutrality point (CNP), the charge carriers in graphene are holes (p-doped) or electrons (n-doped), respectively. (d) Photocurrent comparison at constant illumination, $I_{ph,const}$ (solid lines), and \qty{6}{Hz} AC-light-modulated, $I_{ph,AC}$ (dotted lines), condition. The currents are shown for the same temperatures as in (c). (e) AC-light-modulated responsivity $R_{AC}$, mapped for $V_G$ and a temperature range between 80 to \qty{300}{K}. A dashed line highlights the CNP. All the light measurements were performed at a wavelength $\lambda$ of \qty{1550}{nm} and irradiance ($\mathds{I}_{in}=P_{in}/A$) of \qty{120}{\mu W \per cm^2}. $V_{DS}$ was kept at \qty{1}{V}.}
	\label{fig:ch_TDependence_Figure2}
\end{figure}

One way to specifically introduce gain to a photodetector is the photogating effect. Figure \ref{fig:ch_TDependence_Figure2} (a) shows such an architecture. The \acf{HP} was fabricated by, first, patterning the \acf{CVD} graphene to a channel of $L\times W = 20 \times 1$ \unit{\um ^2}. Then, PbS \acsp{cQD} were spin-coated using the described layer-by-layer approach above. Both the \acs{IFP} and \acs{HP} were placed on the same sample for identical \acs{cQD} film properties (see Supporting Information). The measurements were performed as reported for the \acs{IFP}. Figure \ref{fig:ch_TDependence_Figure2} (b) shows the energy diagram of the graphene-\acs{cQD} interface and the photoresponse of the \acs{HP}. Incoming light is absorbed by the \acs{cQD} layer, generating e-h pairs. The different transfer rates of holes and electrons across the graphene-\acs{cQD} interface result in a predominant hole transfer to graphene. This imbalance renders the \acs{cQD} film negatively charged, photogating the graphene channel. The photogate shifts the charge neutrality point (CNP) of the channel to the right (p-doping), as shown in the $V_G$-dependent $I_{DS}$ curve.

Figure \ref{fig:ch_TDependence_Figure2} (c) shows $I_{DS}$ curves for $V_G$ sweeps at different temperatures measured in the dark. The CNP shifted from 55 to about \qty{45}{V} upon cooling the device from 300 to \qty{80}{K}. The extracted mobilities for the CVD-grown graphene were around \qty{2000}{cm^2 \per Vs} and showed an increase of about 25\% at \qty{80}{K}. The mobility values align with typically reported values for CVD graphene,\cite{Akinwande2019b}, and the temperature dependence might be attributed to phonon scattering induced by the SiO\textsubscript{2} below the graphene channel.\cite{Chen2008}

Figure \ref{fig:ch_TDependence_Figure2} (d) shows the extracted photocurrents at constant illumination and AC-modulated light at \qty{6}{Hz}. The constant illuminated photocurrents were calculated using equation (\ref{eq:ch_TDependence_IphDC}). The photocurrent dips show a sign reversal of the photocurrent due to crossing the CNP of graphene. The double dip of the measurement at \qty{300}{K} might be attributed to hysteresis effects blurring the current traces (see Supporting Information). Hysteresis has previously been found to originate from the charge trapping at the dielectric interface\cite{Wang2010,Nagashio2011,Shulga2016} and in trap sites within the cQD film\cite{Osedach2012,Zhang2015f}. Both sources are present in the \acs{HP} device and might also explain the CNP shift due to different sweeping speeds of the constant and AC-modulated light measurements.

The magnitude of the photocurrent increased moving away from the CNP and then decayed with a $\sim 1/n_e$ (or $\sim 1/n_h$) trend.\cite{Kara2024} $n_e$ and $n_h$ are the charge carrier densities for electrons and holes, respectively. Both the constant and AC-modulated light measurements showed a similar trend in photocurrent but with a difference in the magnitude. The lower magnitude of the AC-modulated measurement can be explained by the slow time response, due to charge trapping in the \acsp{cQD} film, and not allowing the detector enough time to drop to the initial dark current level (see Supporting Information).

$I_{ph}$ reduced with increasing temperature for both constant and AC-modulated illumination. $I_{ph,AC}$ at \qty{210}{K} was the only outlier to the trend. The fact that it was only present in the AC-modulated signal indicates the involvement of charge traps found in those devices.\cite{Grotevent2021,Kara2023} The overall photoresponse improvement for cooling the device down contrasts the \acs{IFP} measurements. The shorter photoexcited charge extraction distance, $\sim 100$ \unit{nm} compared to $\sim 10$ \unit{\um} in the \acs{IFP}, might explain this different behavior. Hence, the reduced mobility in the \acs{cQD} film was not limiting the transfer to graphene. The observed responsivities of $R_{const} \sim 10^5$ \unit{A \per W} reduced about one order in magnitude once the light was modulated. For both cases, a photogain was observed as the $EQE$ values reached $\sim 10^5$ and $10^4$ for constant illumination and AC modulated light, respectively.

Figure \ref{fig:ch_TDependence_Figure2} (e) presents the temperature and gate voltage-dependent AC-modulate responsivity map. The dashed white line indicates the CNP, separating the graphene channel's electron- and hole-doped regions (n/p). Following the photocurrent trend, the responsivity reached $\sim10^4$ \unit{A\per W}, upon cooling the device down to \qty{80}{K}. This one-order-of-magnitude improvement cannot be explained solely by the observed higher mobility in the graphene channel ($\sim1.25$) and follows earlier reports for comparable devices.\cite{Grotevent2021,Kara2023} The photogating leads to about six orders of magnitude higher responsivity values compared to the \acs{IFP}. This is remarkable, taking into account that the \acs{HP} has a photoactive area of \qty{20}{\um ^2} and, thus, about four orders of magnitude fewer photons arrive at the detector for the same illumination.

Although a $R_{AC}$ of $\sim10^4$ \unit{A \per W} was achieved for the \acs{HP}, the higher dark currents in the graphene channel result in elevated noise currents compared to the \acs{IFP}. Based on the measured $1/f-$noise, a specific detective
\begin{equation}
	\label{eq:ch_TDependence_DStar}
	D^* = R \frac{\sqrt{A \Delta f}}{I_{noise}},
\end{equation}
of about $\sim10^9$ Jones could be estimated for the \acs{HP} (see Supporting Information). A crude detectivity estimation for the \acs{IFP}, based on Shot noise $I_{noise} = \sqrt{2eI_{DS,dark} \Delta f}$, results in about $\sim 10^{11}$ \unit{Jones} at \qty{210}{K} ($I_{DS,dark} \approx$ \qty{350}{pA}, Supporting Information). Here $e$ is the elementary charge and $\Delta f$ the frequency bandwidth, taken as \qty{1}{Hz}. The estimated $D^*$ is comparable with commonly reported values in the literature reaching up to $\sim10^{12}$ \unit{Jones}.\cite{Saran2016}

Responsivity $R$ and detectivity $D^*$, although related to each other (equation (\ref{eq:ch_TDependence_DStar})), give very different information regarding the device performance. Where $R$ reveals the magnitude of expected photocurrents, $D^*$ compares the photo- to the noise signal. This can be illustrated if the two detectors compared here would have the same photoactive area of about \qty{20}{\um ^2}. Keeping the bias condition unchanged (constant resistance $\sim L/n_{gaps}W$), would result in the \acsp{IFP}' geometrical dimension of $L \approx$ \qty{115}{nm}, and $W \approx$ \qty{5.775}{\um} ($n_{gaps}$ = 30). With the same illumination condition from above, an unaltered $EQE$ of 3\% is expected, leading to a photocurrent of about \qty{360}{fA} (see Supporting Information). As a lower bound to the noise current, the same bias condition results in the same dark current, which leads to the same noise current of about \qty{10}{fA}. The lowered responsivity thus leads to photocurrents close to the noise limit and would require high-quality amplifying circuits. This is also expressed in a $D^*$-drop to about $10^9$ \unit{Jones}, and yields comparable values to the \acs{HP}. In contrast, however, the immediate photogating of the \acs{HP} gives rise to a responsivity of $\sim10^4$ \unit{A\per W}, and hence 6 orders of magnitude larger photocurrents for the same photoactive area. In addition, the mechanical strength of graphene further allows the fabrication of photodetectors beyond flat \acf{CMOS}-technology as demonstrated on a fiber of \qty{1}{mm} in diameter.\cite{Kara2023}

\section*{Photocurrent Transients}

\begin{figure}[h!tb] 
	\centering
	\includegraphics[scale=0.8]{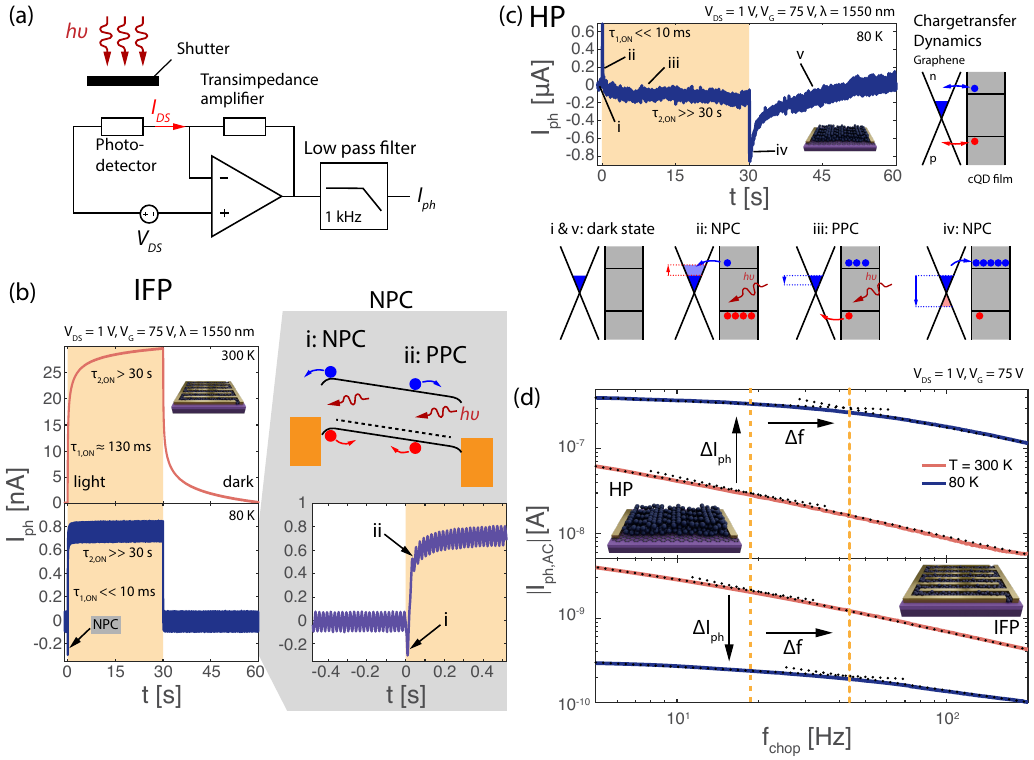}
	\caption{Photocurrent dynamics of \acs{IFP} and \acs{HP}. (a) Transient photocurrent measurement setup. A trans-impedance amplifier was used to apply a source-drain bias $V_{DS}$ and measure the current through the photodetectors. Before subsequent sampling, the signal was low pass filtered at \qty{1}{kHz}. The illumination was controlled by a shutter (\qty{10}{ms} rise/fall time). (b) transient photocurrents for light on (orange shading) and off (white) of the \acs{IFP}, measured at a positive gate voltage and a temperature of 300 (red) or \qty{80}{K} (blue). Two time constants $\tau$ were extracted from the photoresponse upon illumination, where the fast contribution $\tau_{1, on}$ reduces by at least one order of magnitude upon cooling. A \acf{NPC} spike occurs upon illumination of the device at \qty{80}{K}. The enlargement of the region of interest (gray shading) illustrates a formation of a Schottky-like barrier between the \acs{cQD} film and gold electrodes. Illumination at the interface in (i) leads to a fast escape of photoexcited electrons opposing the main current (\acs{NPC}). The illumination of the rest of the detector (ii) leads to a photocurrent contribution in bias direction and re-establishes the \acf{PPC} eventually. The oscillations in the signal are contributions from the line frequency. (c) transient photocurrent of the \acs{HP} showing \acs{NPC} upon illumination and turning off the light. The \acs{NPC} is modeled by a barrier formation at the interface altering the initial band alignment (i). While the photoexcited holes are trapped, a fast electron transfer to graphene renders the cQD-photogate positive and causes an \acs{NPC} spike (ii). The increased hole concentration readjusts the barrier, allowing the cQD film to deplete the holes and reestablishes the \acs{PPC} (iii). Upon turning off the light, electrons transfer back into the \acs{cQD} film, causing another \acs{NPC} spike (iv) before reaching the initial dark current value (v). (d) Photocurrent frequency response (AC-modulated light measured with lock-in technique). \acs{NPC} causing a bandwidth improvement $\Delta f$. This leads to a sacrifice of the photocurrent by $\Delta I_{ph}$ for the \acs{IFP} (bottom). The \acs{HP} (top) shows a simultaneous $\Delta I_{ph}$ and $\Delta f$ improvement.}
	\label{fig:ch_TDependence_Figure3}
\end{figure}

Transient photocurrents were investigated to elaborate on the divergence of constant illuminated and AC-light-modulated photoresponse. Figure \ref{fig:ch_TDependence_Figure3} (a) illustrates the experimental setup. A source-drain voltage was applied to the photodetector, and the current was first amplified, low-pass filtered, and then sampled. An additional gate voltage was applied to the substrate, as described above. A shutter with a rise and fall time of $\sim 10$ \unit{ms} controlled the illumination of the investigated photodetectors.

The transients were measured for positive and negative gate voltages at the specific temperatures relevant to each device (see temperature maps above). Figure \ref{fig:ch_TDependence_Figure3} (b) shows the photoresponse of the \acs{IFP} measured at \qty{300}{K} (red). Upon illumination, the photocurrent rises due to photoexcited charge carriers. An exponential fit reveals the presence of at least two different time constants, where only the fast contribution $\tau_1$ could be estimated. The extracted $\tau_2$ values were comparable or longer than the \qty{30}{s} switching time. All the transients and the extracted time constants can be found in the Supporting Information.

The comparison of the extracted time constants $\tau_1$ can be summarized in three key observations. First, across all measurements, the photocurrent magnitude after \qty{30}{s} was higher for negative gate voltages compared to positive ones. Second, the photoresponse was consistently faster for positive than negative gate voltages. Third, as the devices were cooled down, the time constant decreased only for positive gate voltages.

The photocurrent in a photoconductor depends on the conductivity change upon light illumination, $\Delta \sigma = e(\mu_n \Delta n + \mu_p \Delta p)$. $\Delta n$ and $\Delta p$ are the light-induced charge carrier density changes for electrons and holes. Similarly, $\mu_n$ and $\mu_p$ describe the electron and hole mobilities, respectively. The influence of the gate voltage on the photocurrent, thus, needs to be explained by considering the modulation of charge carrier density and mobility.

Although incoming photons initially generate the same number of charge carriers for electrons and holes ($\Delta n = \Delta p$), the gate voltage can tune their recombination rate before they are extracted to the electrodes. The photogenerated holes (majority carriers) can flow easily in the p-type channel for negative gate voltages. In contrast, the photogenerated electrons (minority carriers) will be suppressed and readily recombine with excess holes. In the other extreme for positive gate voltages, the holes are first depleted before an inversion channel of electrons forms between the source and drain contacts. In that case, the photogenerated holes recombine within the inversion layer before reaching the electrodes, whereas the photogenerated electrons can travel more easily through the channel instead. However, because the holes are the majority carriers in the first place, the overall recombination rate is higher for electrons. This results in a larger photocurrent magnitude in the case of negative than positive gate voltages.

Considering that the mobility is related to the response speed, $\tau_1$ can be assigned to the mobility of the photogenerated holes ($\mu_p$) and electrons ($\mu_n$) for positive and negative gate voltages, respectively. The consistently faster photoresponse for positive gate voltages thus suggests $\mu_n>\mu_p$. This effect becomes particularly pronounced when cooling the device, where $\tau_1$ decreases by more than one order of magnitude for positive gate voltages. In agreement with a complete depletion of the channel only upon cooling the device (see Figure \ref{fig:ch_TDependence_Figure1} (c)), the formation of an inversion layer allowing electrons to propagate more easily is very likely. On the contrary, cooling changes $\tau_1$ only slightly for negative gate voltages as the holes (majority carrier) still carry the photoresponse.

Faster mobilities for electrons in \acs{EDT} treated PbS \acs{cQD} films were previously attributed to favorably aligned transport across trap states within the energy gap of the \acs{cQD} film.\cite{Nagpal2011,Zhang2015b} For a positive gate voltage of $V_G = 75$ \unit{V} (n-type inversion layer), an activation energy of \qty{40}{meV} could be estimated from the temperature-dependent photocurrent measurements (Supporting Information). Taking reported trap state energies for the electrons around 100 to \qty{300}{meV} into account,\cite{Konstantatos2007a,Konstantatos2008}, would thus support electron transport across formed trap states.

Consequently, the discrepancy between constant and AC light modulation can be explained by considering the difference in hole and electron photoresponse. Under modulated AC light, the faster electron dynamics at positive gate voltages dominate the photoresponse, whereas constant illumination at negative gate voltages highlight the slower but higher hole-driven photocurrent. This effect becomes more pronounced upon cooling, as depletion of the p-channel facilitates the formation of an inversion layer.

Figure \ref{fig:ch_TDependence_Figure3} (b) also shows the transient photocurrent at \qty{80}{K} and a positive gate voltage of \qty{75}{V} (blue). Upon illumination, a \acf{NPC} spike renders the device more resistive before a \acf{PPC} is re-established. \acp{NPC} are attributed to trion formation, photothermal heating, absorption and desorption of molecules, light-activated trap states, and interfaces.\cite{Tailor2022} The commodity of most observation is the presence of charge carrier trapping sites. For example, in a Bi-doped organic-inorganic halide perovskite structure, the negative photocurrent was attributed to the light-induced formation of a deep trapping site, enhancing recombination rates.\cite{Haque2019} In GeS nanowires, the observed \acs{NPC} was assigned to the trapping of photoexcited holes at the interface.\cite{Zhao2021} Figure \ref{fig:ch_TDependence_Figure3} (b) proposes such a possible Schottky-like barrier formation at the interface leading to trapping of photoexcited holes. Upon absorption of a photon next to the interface (i), electrons can quickly escape to the contact, whereas the holes are slowed down, giving rise to the observed \acs{NPC}. Photogenerated charge carriers in other parts of the film (ii) still follow the applied bias voltage, eventually causing the \acs{PPC} to take over.

Figure \ref{fig:ch_TDependence_Figure3} (c) shows the transient photocurrent for the \acs{HP} device investigated also at positive gate voltage and \qty{80}{K}. The photogating amplifies the described \acs{NPC} effect, rendering a fast, $\tau_{1,on}<<10$ \unit{ms}, photocurrent of about \qty{0.6}{\micro A}. The resulting $EQE\sim10^4$ proofs the involvement of gain by photogating. All \acsp{NPC} were observed only at \qty{80}{K} and for positive gate voltages.

Charge transfer at the graphene-\acs{cQD} interface is bidirectional and depends on the energetic band alignment between the two materials. A barrier at the interface can alter the initial band alignment (i), leading to the trapping of photoexcited holes. This enables the transfer of faster-moving electrons to graphene (ii) and causes an \acs{NPC} spike upon initial light illumination due to positive photogating. Eventually, the increased hole concentration in the cQD film rearranges the barrier and leads to a favorable hole transfer to graphene. This, in contrast, recovers the steady-state \acs{PPC} in (iii). The described process repeats upon turning off the light in (iv), where the faster-moving electrons transfer back into the \acs{cQD} film, rendering the photogate more negative and causing another \acs{NPC} spike before reaching the dark current value in the channel (v).

The \acs{NPC} effect, reflected in photocurrent spikes, is of particular interest for the \acs{HP}. The initial gain in this device architecture arises from long lifetimes of photoexcited e-h pairs, compared to fast charge transport within the graphene channel. While this is highly desirable to resolving low light levels down \unit{nW\per cm^2},\cite{Konstantatos2012,Goossens2017} the extended lifetimes conversely slow down the photoresponse speed. One way to increase the detector's frequency response is by pulsing the back gate to drain the trapped charges in the \acs{cQD} film.\cite{Konstantatos2012} The observed \acs{NPC} here has a similar effect through light-induced draining of those trapped charges.

In Figure \ref{fig:ch_TDependence_Figure3} (d), the frequency response of both \acs{IFP} and \acs{HP} are presented up to \qty{200}{Hz}. The frequency response was acquired using the described lock-in technique above. When the \acs{IFP} is cooled to \qty{80}{K}, the bandwidth, $\Delta f$, increases by about \qty{25}{Hz}. This is accompanied by a loss in photocurrent, $\Delta I_{ph}$, of about one order of magnitude. On the contrary, the \acs{HP} device shows a simultaneous increase by $\Delta f$ and an order of magnitude rise by $\Delta I_{ph}$. As the photoresponse does not immediately cease, frequencies up to \qty{100}{kHz} could be verified (Supporting Information).

\section*{Conclusion}

Here, we compared photoconductors, \acs{IFP}, (area of \qty{7500}{\um ^2}) and hybrid phototransistors, \acs{HP}, (area of \qty{20}{\um ^2}) fabricated with identical \acs{EDT} treated PbS \acs{cQD} film (on same sample). The \acs{HP} architecture demonstrated a responsivity $R$ of $10^4$ \unit{A \per W} (\qty{6}{Hz}, wavelength \qty{1550}{nm}), which is six orders of magnitude large than that of the \acs{IFP}. A detectivity $D^*$ of $10^{11}$ \unit{Jones} was estimated for the \acs{IFP}. However, reducing the 4-orders of magnitude larger photoactive area of the \acs{IFP} to a technologically relevant dimension of \qty{20}{\um ^2} reduced its $D^*$ to about $10^9$ \unit{Jones}. This is the same order of magnitude as the detectivity found for the \acs{HP}. The inherent photogain in \acs{HP} results, thus, in a readily accessible photosignal of \unit{nA} at the same signal-to-noise ratio ($D^*$). In addition, the mismatch of the photoresponse between AC-modulated and constant light illumination was explained by the faster electron dynamics compared to the slower but higher hole-driven photocurrent. Finally, the temperature and gate voltage-activated photocurrent dynamics led to an observed \acf{NPC} causing a simultaneous bandwidth and photocurrent improvement at \qty{80}{K}. This higher frequency response enabled the operation of the \acs{HP} up to \qty{100}{kHz}.

\FloatBarrier

\section*{Methods}

\subsection*{Device Fabrication}
The fabrication of the investigated devices was described in \cite{Kara2024} but briefly, Ti/Au (\qty{5}{nm}/\qty{40}{nm}) contacts were photolithographically patterned on a p-Si with \qty{285}{nm} chlorinated dry thermal SiO\textsubscript{2} by e-beam evaporation. CVD graphene was grown on a commercial Cu-foil (Alfa Aesar Foil purchased in 2022, No. 46365). Prior to graphene growth, the Cu-foil was prepared by slightly etching for 1 min in Cu-etch (APS-100) and electropolishing (70\% H\textsubscript{3}PO\textsubscript{4}). Then the Cu-foil was cleaned in acetone, IPA, and deionized water. The Cu-foil was then transferred to HNO\textsubscript{3}, deionized water, immersed in ethanol, and dried with N\textsubscript{2}. Then the Cu-foil was annealed at 1000 $^\circ$C under a mixture of H\textsubscript{2} (20 sccm) and Ar (200 sccm) at 1 mbar for 75 min. Subsequently, the graphene growth was initiated by introducing CH\textsubscript{4} at 0.05 sccm for 25 min and increasing the flow to 0.2 sccm for an additional 10 min. 50K PMMA resist (Allresist) was spin-coated onto the graphene/copper surface, and graphene on the backside of the foil was etched by reactive ion etching. Then the Cu was etched in a commercial Transene Cu-etchant, the floating graphene was placed in deionized water, and wet transferred on the pre-patterned substrate. A Cu mask (\qty{100}{nm}) was prepared with a liftoff process by an e-beam lithographically defined structure. Graphene channels were O\textsubscript{2}-plasma etched, and Cr/Au top-contacts (2/40 \unit{nm}) were evaporated.

\subsection*{PbS \acs{cQD} Film Fabrication}
For synthesizing PbS QDs, the procedure introduced by Hines et al. \cite{Hines2003} was slight adapted as described previously\cite{Kara2023}. The absorbance spectrum was acquired with by UV-Vis spectroscopy (Jasco V-670). For that, the PbS \acsp{cQD} were dispersed in tetrachloroethylene. A conductive film of PbS \acs{cQD} was prepared by \acf{LbL} spin-coating at ambient. PbS \acsp{cQD} in octane (\qty{20}{mg \per ml}) were spin-coated (\qty{2500}{rpm}, \qty{45}{s}), and a drop of 2 vol\% \acf{EDT} in acetonitrile was placed for \qty{30}{s} subsequently. After spinning the sample dry, a drop of acetonitrile and one drop of octane were added while spinning (\qty{2500}{rpm}, \qty{45}{s}). The procedure was repeated six times, resulting in a $\sim 170$ \unit{nm} thickness.

\subsection*{Device Characterization}
The devices were characterized by modulating a broadband light source (Thorlabs, SLS201) with a chopper (Thorlabs, MC2000B-EC). Monochromatic light (Princeton Instruments, SpectraPro HRS-300 spectrometer with grating 150 G/mm, blaze 0.8 $\mu$m and long-pass filters: 400, 600, 800, 1200, 1900 nm) was collimated with a lens and split with a 50/50 Polkadot beamsplitter and aligned to onto the sample, that was placed into an optically accessible cryostat (JANIS ST-100) with a quartz glass window. The detector (Gentec, UM-9B-L) measured the reference beam using a lock-in amplifier (Stanford Research System, SR865A). The gate voltage and source-drain bias were applied with SMUs (Keithley, 2614B and 2450). The photovoltage was measured over a shunt resistance of \qty{1}{k\Omega} with a lock-in amplifier (Stanford Research System, SR860).

The noise of the \acf{HP} was characterized with a battery-powered trans-impedance amplifier (Stanford Research Systems, SR570) by applying a source-drain voltage bias and measuring the current. The signal was low pass filtered (10 kHz) and sampled at \qty{500}{kHz} with a data acquisition board (National Instruments, USB6341). The DC offset was removed for the power spectral densities estimates, and $10\times$ one-second-long time traces were averaged. The sample was measured in the dark.

For the photocurrent transients, a high-speed shutter (Thorlabs, SHB1T) and line-powered trans-impedance amplifier (Stanford Research Systems, SR570) were used to apply a source-drain bias voltage and measure the current. The signal was low-pass filtered at \qty{1}{kHz}, sampled at \qty{500}{kHz} and subsequently downsampled to \qty{10}{kHz}. A gate voltage was applied by an SMU (Keithley, 2614B). 

All measurements were performed in vacuum (4$\cdot 10^{-7}$ \unit{mbar}), and the device was cooled down to \qty{80}{K} by liquid N\textsubscript{2}. The external quantum efficiency ($EQE$) was calculated using $EQE=RE_{ph}/e$.

\begin{acknowledgement}

The authors thank FIRST-Lab (Center for Micro- and Nanoscience) at ETH Zurich for access to the clean-room and the Swiss National Science Foundation (SNSF, project no. 200021 182790) for financial support.

\end{acknowledgement}

\begin{suppinfo}

Additional information and experimental data to PbS \acs{cQD} absorption spectra, spectral photoresponse of the photodetectors, AC and constant illuminated photoresponses, activation energy extraction, temperature dependent transfer characteristics, photoresponse estimation of photodetector with downscaled dimension, transient photocurrents with extracted rise/fall times, frequency dependent photocurrent measurements, noise power spectral density, IVs can be found in the supporting information. Supporting information is available free of charge via the internet at http://pubs.acs.org.

\end{suppinfo}

\section*{Author Contribution}

G.K. and I.S. conceived the study and planned the experiments. G.K. fabricated the samples, performed the measurements and analyzed the data. L.F. performed additional high frequency measurements. D.D. synthesized PbS colloidal QDs. R.F. grew graphene. G.K., I.S. and M.C. discussed the data. I.S., M.C., and M.K initiated and supervised the project. G.K. wrote the manuscript with inputs from and discussions with all authors.


\newpage
\providecommand{\latin}[1]{#1}
\makeatletter
\providecommand{\doi}
  {\begingroup\let\do\@makeother\dospecials
  \catcode`\{=1 \catcode`\}=2 \doi@aux}
\providecommand{\doi@aux}[1]{\endgroup\texttt{#1}}
\makeatother
\providecommand*\mcitethebibliography{\thebibliography}
\csname @ifundefined\endcsname{endmcitethebibliography}
  {\let\endmcitethebibliography\endthebibliography}{}


\end{document}